\documentclass[superscriptaddress,twocolumn,secnumarabic,
amssymb,amsmath,nobibnotes,aps,prd,showpacs,nofootinbib]{revtex4}%
\usepackage{graphicx}
\usepackage{epsf}
\usepackage{bm}
\usepackage{amsmath}
\usepackage{amsfonts}
\usepackage{amssymb}
\usepackage{epstopdf}
\usepackage{natbib}
\usepackage{color}%
\providecommand{\U}[1]{\protect\rule{.1in}{.1in}}
\newcommand{\be}{\begin{equation}}
\newcommand{\ee}{\end{equation}}

\newcommand{\mincir}{\raise
-3.truept\hbox{\rlap{\hbox{$\sim$}}\raise4.truept\hbox{$<$}\ }}
\newcommand{\magcir}{\raise
-3.truept\hbox{\rlap{\hbox{$\sim$}}\raise4.truept\hbox{$>$}\ }}

\usepackage{color}
\begin{document}


\title{Constraining a dark matter and dark energy interaction scenario with a dynamical equation of state}

\author{Weiqiang Yang}
\email{d11102004@mail.dlut.edu.cn}
\affiliation{Department of Physics, Liaoning Normal University, Dalian, 116029, P. R. China}

\author{Narayan Banerjee}
\email{narayan@iiserkol.ac.in}
\affiliation{Department of Physical Sciences, Indian Institute of Science Education and Research, Kolkata, Mohanpur$-$741246, West Bengal, India}

\author{Supriya Pan}
\email{span@iiserkol.ac.in}
\affiliation{Department of Physical Sciences, Indian Institute of Science Education and Research, Kolkata, Mohanpur$-$741246, West Bengal, India}


\begin{abstract}
In this work we have used the recent cosmic chronometers data along with the latest estimation of the local Hubble parameter value, $H_0$ at 2.4\% precision as well as the standard dark energy probes, such as the Supernovae Type Ia, baryon acoustic oscillation distance measurements, and cosmic microwave background measurements (PlanckTT $+$ lowP) to constrain a dark energy model where the dark energy is allowed to interact with the dark matter. A general equation of state of dark energy parametrized by a dimensionless parameter `$\beta$' is utilized. From our analysis, we find that the interaction is compatible with zero within the 1$\sigma$ confidence limit.
We also show that the same evolution history can be reproduced by a small pressure of the dark matter.
\end{abstract}

\maketitle

\section{Introduction}

According to the present  observations, dark energy and dark matter are two chief constituents of our universe comprising respectively about 69\% and 26\% of its total energy \cite{Ade:2015xua}. The dark energy (DE) is the one which has an effective negative pressure (the equation of state parameter $w_{DE}=\frac{p_{DE}}{\rho_{DE}} < -\frac{1}{3}$ ) and drives the alleged late accelerated expansion of the universe while the pressureless dark matter component is responsible for the structure formation of our universe. The remaining 5\% of the total energy of our universe is contributed by the baryonic matter. Now, to portray the cosmic evolution one can construct several cosmological models with the available observational data from different independent sources. Amongst them the current observations favor the so-called $\Lambda$CDM cosmology, in which the cosmological constant $\Lambda$ acts as a dark energy. However,
$\Lambda$CDM cosmology encounters a big problem, the ``cosmological constant problem'' \cite{Carroll:2000fy}, which is related to the huge descrepancy between the theoretically predicted and the observationally required values. So, an alternative to $\Lambda$CDM cosmology, namely the $w$CDM cosmology \cite{AT2010} was suggested in which the equation of state parameter for the dark energy $w$  ($\neq -1$), could be either a constant or evolving with the cosmic time while both DE and CDM are conserved separately.  But, this has the cosmic coincidence problem \cite{Zlatev:1998tr} which questions ``why the energy densities of dark energy and matter are of the same order while they evolve completely differently with the evolution of the universe?''. We note that $\Lambda$CDM cosmology can not escape from the coincidence problem too. It was suggested that a mutual interaction between dark matter and dark energy can potentially resolve this issue. This would mean that the total energy of dark matter plus dark energy is conserved unlike their separate conservation as in $\Lambda$CDM or $w$CDM cosmology. The idea of an interacting dark sector raised a lot of interest for several reasons. The inclusion of interaction provides richer and wider possibilities and reduces to the usual non-interacting cosmology under the low interaction limit. However, recently, a large number of investigations argue that the current observational data prefer a nonzero interaction in the dark sector \cite{Salvatelli:2014zta,Yang:2014hea,Yang:2014gza,yang:2014vza,Nunes:2016dlj,Kumar:2016zpg,Yang:2016evp,Sola:2016ecz, vandeBruck:2016hpz}.  For a review of the interacting models, we refer to \cite{DE_DM_1, DE_DM_2}). \\

In the present work we have considered  a scenario where the cold dark matter and the dark energy are interacting with each other. Specifically, we have considered the spatially flat Friedmann-Lema{\^\i}tre-Robertson-Walker (FLRW) universe where both CDM and DE have barotropic equations of state, and in addition to that, the equation of state for DE is varying with the cosmic evolution. We have considered a general equation of state for DE whose nature is characterized by a dimensionless real parameter $\beta$. For $\beta = 1$, $-1$, one recovers the Chevallier-Polarski-Linder (CPL) \cite{cpl1,cpl2} and linear parametrizations \cite{lin1,lin2,lin3} respectively, on the other hand for $\beta \longrightarrow 0$, one recovers the logarithmic parametrization \cite{log}. For any values of $\beta$ other than ($-1, 0, 1$), we have a wide variety of dark energy parametrizations. Our aim is to constrain the interaction between CDM and DE where the later component, i.e., DE, has a dynamical equation of state, with the use of the recently released cosmic chronmeters data, local value of the Hubble parameter estimation as well as with other standard dark energy probes. \\

We find that the interaction, if there is any, is indeed very small and a non-interacting model is always within the $1\sigma$ confidence region in the parameter space. \\

In the presence of the interaction, the dark matter will have a departure from the standard `$\frac{1}{a^{3}}$' form. This departure can in fact be the result of the presence of a small pressure of the dark matter in a non-interacting scenario. We also discuss this possibility with the same observational data sets. The constraints on various cosmological parameters are very close in these two descriptions. The major qualitative difference is in the evolution of the matter density. The density of the cold dark matter redshifts faster than the standard $\frac{1}{a^{3}}$ rate in the non-interacting case, whereas it redshifts slower than the standard CDM in the interacting scenario.  \\

The paper is organized as follows. In section \ref{ide-equations} we briefly describe the background equations. Section \ref{sec-models} introduces the models on variable equation of state for DE. Section \ref{sec:data} briefly describes the observational data sets employed in our analysis. In section \ref{sec:results} we present the results of the analysis. In section \ref{pressure}, we discuss the possibility of replacing the interaction by a small pressure of the dark matter. Finally, in section \ref{sec:summary} we include a discussion on the results obtained.

\section{Interacting dark energy}
\label{ide-equations}

We consider that the geometry of our universe is described by a spatially flat FLRW universe whose line element is given by

\begin{equation}
\label{metric}
 {\rm d}s^2= -{\rm d}t^2+ a^2(t)\Bigl[  {\rm d}r^2+ r^2 \left({\rm d}\theta^2 + \sin^2 \theta \, {\rm d}\phi^2\right) \Bigr   ],
\end{equation}
where $a(t)$ is the  scale factor of the universe. Now, in such a background, one can write down the Einstein's field equations as

\begin{eqnarray}
\label{fld-eq}
H^2= \frac{8\pi G}{3}\, \left(\rho_r+ \rho_b+ \rho_{dm} + \rho_{DE}  \right),\\
2 \dot{H}+ 3 H^2= - 8 \pi G (p_r+ p_b+ p_{dm}+p_{DE}),
\end{eqnarray}
in which an overhead dot represents a derivative with respect to the cosmic time `$t$', $H= \dot{a}/a$ is the Hubble parameter, $\rho_{r}$, $\rho_b$, $\rho_{dm}$, and $\rho_{DE}$ are respectively the energy densities of radiation, baryons, cold dark matter (CDM) and dark energy (DE) whereas $p_r$, $p_b$, $p_{dm}$, $p_{DE}$ stand for the corresponding pressures. Now we consider a scenario where CDM and DE are coupled. Since the energy densities of baryon and radiation are insignificant in comparison with the other two components, we assume that they are conserved separately, which indicates that $\rho_b \propto a^{-3}$ and $\rho_r \propto a^{-4}$. \\

In what follows, we shall assume that DE and CDM interact with each other and thus are not conserved by themselves, they do so only in tandem. Furthermore, it is also assumed that the DE formally mimics a fluid, and its energy momentum tensor can be written in the fashion as one does for a fluid. In the presence of this interaction, using the total  conservation equation $u_{\mu}T^{\mu \nu}_{;\nu} = 0$, in which $T^{\mu \nu}_{;\nu}$ denotes the total energy-momentum tensor of the dark fluids defined by
$T^{\mu \nu}= T^{\mu \nu}_{dm} + T^{\alpha \beta}_{DE}$, one can write,
\begin{equation}\label{cons-eqn}
\dot{\rho}_{dm} + 3\frac{\dot{a}}{a}\rho_{dm} = -\dot{\rho}_{DE} -
3\frac{\dot{a}}{a}\left(\rho_{DE} +
p_{DE}\right)=Q,
\end{equation}
where $Q$ is the interaction between DE and CDM. The pressure of the fluid representing the CDM is chosen to be zero as usual. \\

There is no general form of the inetraction $Q$, vaious forms are chosen as an ansatz. Typical choices found in the literature are like $Q \propto \rho_{dm}$, $Q \propto \rho_{DE}$ or $Q\propto (\rho_{dm}+ \rho_{DE})$, or even more complicated forms. There have been an extensive analysis with different phenomenological interactions in the last couple of years \cite{Wetterich:1994bg, Amendola:1999er, Billyard:2000bh, Olivares:2005tb,He:2008tn,delCampo:2008jx,delCampo:2008sr,Chimento:2009hj, Quartin:2008px,Valiviita:2009nu, Clemson:2011an, Pan:2013rha, Faraoni:2014vra, Pan:2014afa, Chen:2011cy, Tamanini:2015iia, Pan:2012ki,Duniya:2015nva, Valiviita:2015dfa, Marcondes:2016reb, Mukherjee:2016lor, Pan:2016ngu, Mukherjee:2016shl,Sharov:2017iue}. However, another possibility with which one may model the effect of interaction is to choose a small deviation, of the evlution of the CDM,  from its standard evolution with the scale factor as $\rho_{dm}\propto a^{-3}$. That means, one may consider that due to presence of interaction between these sectors, cold dark matter evolves as \cite{Nunes:2016dlj,Kumar:2016zpg,Wang:2004cp,Costa:2007sq,Nunes:2014qoa}

\begin{equation}\label{cdm-evolution}
\rho_{dm} = \rho_{dm,0}\, a^{-3+\delta}\, ,
\end{equation}
where $\rho_{dm,0}$ is the present value of $\rho_{dm}$ and $\delta$ indicates the interaction between DE and CDM. In general, $\delta$ could be either constant or time dependent, but its value should be small as the evolution of the CDM cannot deviate much from the standard  $\rho_{dm}\propto a^{-3}$. In our investigation we consider $\delta$ to be varying very slowly with the evolution of the universe, such that its variation over a time less than the Hubble scale can be safely neglected, i.e., $\dot{\delta} \sim 0$. \\

It is readily seen that $\delta = 0$ is the non-interacting scenario. Furthermore, one can understand the direction of the flow of energy by the sign of $\delta$. If $\delta < 0$, then with the increase of the scale factor, $\rho_{dm}$ decreases more  rapidly than its standard redshift without any interaction, hence, one can identify that due to presence of interaction the energy flow occurs from CDM to DE. Follwoing the same line, $\delta > 0$ indicates the energy flow from DE to CDM. We now assume that the dark energy satisfies an equation of state $w_{DE}= p_{DE}/\rho_{DE}$, where $w_{DE}$ is identified as the equation of state parameter for DE. For a constant $w_{DE}$, a recent analysis can be found in \cite{Nunes:2016dlj, Kumar:2016zpg}. In particular, in \cite{Nunes:2016dlj} it was shown that an interaction is mildly favored by the combined analysis of several observational data and the equation of state $w_{DE}$ could be of a marginally phantom character, that means $w_{DE} < -1$. On the other hand, Kumar and Nunes \cite{Kumar:2016zpg} showed that the same interacting scenario with constant $w_{DE}$, in presence of massive netrinos, indicates a small interaction but the equation of state parameter $w_{DE}$ exhibits a quintessential behavior, i.e. $w_{DE} > -1$, which is different from \cite{Nunes:2016dlj}. This difference might be due to the massive neutrinos in the background evolution. In the present work we will concentrate on the possible evolution scenario when $w_{DE}$ is variable. Now, inserting (\ref{cdm-evolution}) into (\ref{cons-eqn}) one arrives at the following first order differential equation,

\begin{equation}
\frac{d\rho_{DE}}{da}+ \frac{3}{a}\,(1+w_{DE})\rho_{DE}= -\,\delta\, \rho_{dm,0}\,a^{-4+ \delta},
\end{equation}
and consequntly one can find the evolution of $\rho_{DE}$ as

\begin{equation}\label{DE-evolution}
\rho_{DE} = \frac{(1+z)^3}{f(z)}\, \left[\rho_{DE,0} f(0)+ \delta\, \rho_{dm,0}\int_{0}^{z} (1+z)^{2-\delta} f(z) dz\right],
\end{equation}
where $\rho_{DE,0}$ is the present value of $\rho_{DE}$, $f(0)$ is the value of $f(z)$ at $z= 0$ and $f(z)$ is given by

\begin{equation}\label{fz}
f(z)= \exp\left(-3 \int \frac{w_{DE}}{1+z} dz \right).
\end{equation}

Thus, it is clear from equation (\ref{DE-evolution}) that for any given $w_{DE}$ as a function of $z$, DE evolution can be found out. Let us remark that using the evolution of CDM as in (\ref{cdm-evolution}) into the balance equation for CDM in (\ref{cons-eqn}), one finds the evolution of the interaction $Q = \delta H \rho_{dm,0}\; a^{-3+\delta} =\delta H \rho_{dm} $, which means that $\delta$ enhances the interaction between the dark sector.

Now, from the first equation of (\ref{fld-eq}) it is easy to write

\begin{align}
\left(\frac{H}{H_0}\right)^2= \Omega_{r0}(1+z)^4+ \Omega_{b0}(1+z)^3+ \Omega_{dm,0}(1+z)^{3-\delta}\nonumber\\+ \frac{(1+z)^3}{f(z)}\, \left[\Omega_{DE,0}\, f(0)+ \delta\, \Omega_{dm,0}\int_{0}^{z} (1+z)^{2-\delta} f(z) dz\right],
\end{align}
where $\Omega_{r0}+\Omega_{b0}+\Omega_{dm,0}+\Omega_{DE,0}= 1$. \\
A subscript $0$ indicates the present value of the quanity.

\section{Evolving dark energy models}
\label{sec-models}

Parametrization of the equation of state of the dark energy may be considered in various ways. In fact, three different parametrizations are most commonly used, namely (i) Chavallier-Polarski-Linder (CPL) \cite{cpl1,cpl2}, (ii) linear \cite{lin1, lin2, lin3} and (iii) logarithmic \cite{log} paramerizations, described respectively as

\begin{eqnarray}
&&w_{DE} (z) = w_0 + w_{1} \frac{z}{1+z}\, ,\label{w-cpl}
\end{eqnarray}
\begin{eqnarray}
&&w_{DE} (z) = w_0 + w_{2}\, z,\label{w-linear}
\end{eqnarray}
\begin{eqnarray}
&&w_{DE} (z) = w_0 + w_{3}\, \ln(1+z),\label{w-log}
\end{eqnarray}
where $w_{1}$, $w_{2}$, $w_{3}$ are real constants, and $w_0$ is the present value of $w_{DE}$. \\

However, it is interesting to note that the above three dark energy parametrizations can be written in a single parametric way characterized by a single parameter `$\beta$' which yields all the three as special cases. The general equation is given by

\begin{equation}\label{model1}
w_{DE} (z)= w_0 -w_{\beta} \left[\frac{(1+z)^{-\beta}-1}{\beta}\right],
\end{equation}
where $w_0$ is the present values of $w_{DE}$ and $w_{\beta}$ is a real parameter. The parametrization in eqn. (\ref{model1}) was introduced by Barboza \textit{et al} \cite{Barboza:2009ks}. Now, it is easy to see that for  $\beta = 1$, CPL parametrization \cite{cpl1,cpl2} is recovered. Similarly, one can find the linear parametrization \cite{lin1, lin2, lin3} for $\beta = -1$. Finally, in the limiting case $\beta \longrightarrow 0$, one easily finds the logarithmic parametrization \cite{log}. \\

\begin{center}
\begin{figure}[!htbp]
	\includegraphics[width=8.0cm,height=6.2cm]{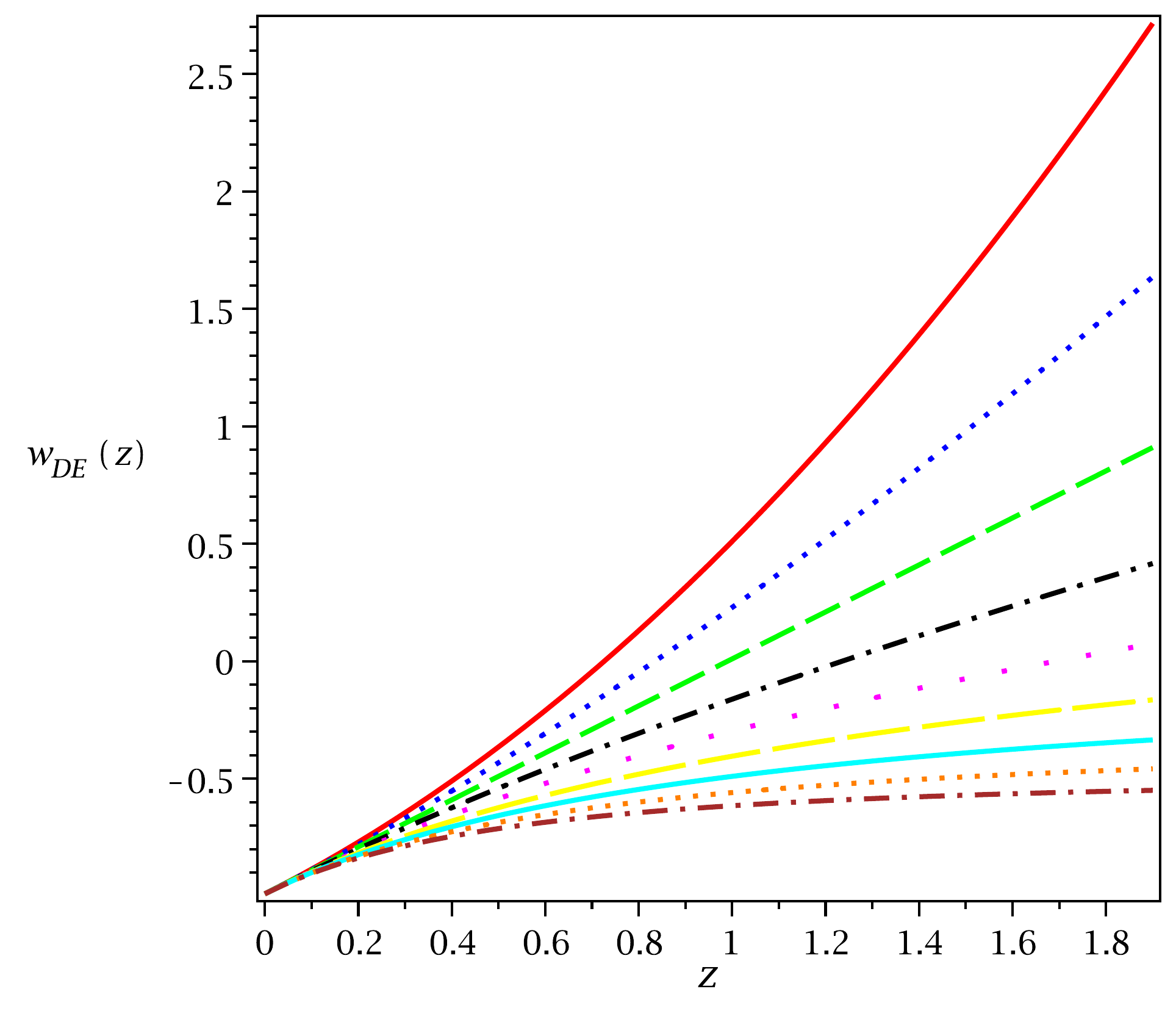}
	\caption{Qualitative evolution of the dark energy equation of state $w_{DE} (z)$ in Eq.  (\ref{model1}) has been shown for $w_{\beta} = 1$,  with different values of the $\beta$ parameter such as $\beta =$ $-2, -1.5, -1, -0.5,~ 0.001,~ 0.5,~ 1,~ 1.5,~ 2$. The lowest plot is for $\beta = 2$, and as $\beta$ decreases, the plots go up in the anticlockwise direction.}
	\label{EoS-DE-plot}
\end{figure}
\end{center}

Other values of $\beta$ describes a very wide range of parametrization of the dark energy equation of state. However, in the current work we shall consider a single general model, namely that given by (\ref{model1}) and check how $\beta$ is constrained, along with other cosmological parameters by the observational data sets. We note that for the single parametrization given in eqn. (\ref{model1}), $f(0) = 1$ where $f(z)$ can be found in eqn. (\ref{fz}). 
In Figure \ref{EoS-DE-plot} we describe the qualitative evolution of $w_{DE} (z)$ for  $w_{\beta}> 0$ as with different values of the $\beta$ parameter such as $-2, -1.5, -1, -0.5,~0.001,~ 0.5,~ 1,~ 1.5,~ 2$. We fix $w_{\beta} =1$.  Since the current value of the dark energy equation of state is very close to $-1$, we take $w_0 = -0.99$. We mention here that there is no significant qualitative change in the nature of the plots will be seen for different values of $w_0$, and also for small positive values of $w_{\beta}$.

\section{The Observational Data}
\label{sec:data}

We shall briefly discuss the data sets that are used in the present work in order to constrain the parameters in a scenario where the dark sectors interact amongst themselves.

\subsection{Cosmic chronometers data}
\label{cc-data}

The cosmic chronometer approach is a method to determine the Hubble parameter values at different redshifts with the use of most massive and passively evolving galaxies in our universe. These galaxies are known as cosmic chronometers (CC). The idea is to determine $dz/dt$ and hence the Hubble parameter $H(z)= -\left(1/1+z \right)dz/dt$. Since the measurement of $dz$ is obtained through spectroscopic method with high accuracy therefore a precise measurement of the Hubble parameter lies on the precise measurement of the differential age evolution $dt$ of such galaxies, and hence these measurements are considered to be model independent. A detailed description  about cosmic chronometer method can be found in \cite{Moresco:2016mzx}. Here, we use $30$ measurements of the Hubble parameter in the redshift interval $0 < z< 2$  \cite{Moresco:2016mzx}.

\subsection{Local value of the Hubble parameter ($H_0$)}

Along with the cosmic chronometers data, we include the local value of the Hubble parameter which yields $H_0= 73.02 \pm 1.79$  km/s/Mpc with 2.4\% precision as reported in \cite{Riess:2016jrr}.

\begin{center}
	\begin{table}
		\begin{tabular}{cccccccc}
			\hline\hline
			Parameters & Priors & Mean with errors & Best fit \\ \hline
			$\Omega_b h^2$ & $[0.005, 1]$ &$    0.02237_{-    0.00034- 0.00056-  0.00071}^{+ 0.00027+ 0.00059+ 0.00076}$ & $ 0.02240$\\
			$\Omega_{dm} h^2$ & $[0.01, 0.99]$ & $    0.1227_{-    0.0036- 0.0065-    0.0101}^{+ 0.0035+  0.0071 + 0.0082}$ & $ 0.1259$\\
			$100\theta_{MC}$ & $[0.5, 10]$ & $ 1.04073_{-    0.00044- 0.00089- 0.00117}^{+ 0.00044+ 0.00088+ 0.00128}$ & $    1.04064$\\
			$\tau$ &  $[0.01, 0.8]$ &$0.075_{- 0.018-    0.035- 0.046}^{+ 0.018+ 0.037+ 0.048}$ & $  0.066$\\
			$n_s$ & $[0.5, 1.5]$ &$    0.9625_{- 0.0063-    0.0107 - 0.0140}^{+    0.0056 + 0.0113 + 0.0149}$  & $    0.9574$\\
			${\rm{ln}}(10^{10} A_s)$  & $[2.4, 4]$ & $    3.085_{- 0.035- 0.071-  0.089}^{+ 0.036 + 0.069+ 0.095}$ & $ 3.071$\\
			$\delta$ & $[-2, 2]$ &$    0.00214_{- 0.00300- 0.00507- 0.00761}^{+    0.00280+ 0.00542+ 0.00749}$ & $    0.00378$\\
			$w_0$ & $[-2, 0]$ &$  -1.030_{- 0.153-    0.244-  0.308}^{+ 0.114+ 0.269+  0.334}$ & $   -0.944$\\
			$w_{\beta}$ & $[-3, 3]$ & $  -0.218_{- 0.267-  0.870-  0.999}^{+    0.485+    0.640+    0.795}$ & $   -0.419$\\
			$\beta$ & $[-3, 3]$ & $    0.960_{- 0.367-  1.113-    1.636}^{+    0.546+    1.100+    1.154}$ & $    0.350$\\
			$\Omega_{m0}$ &  $-$ & $    0.307_{-    0.012-    0.022- 0.029}^{+    0.011+    0.023+  0.030}$ & $  0.321$\\
			$\sigma_8$ &  $-$ & $  0.847_{- 0.019-    0.037- 0.048}^{+ 0.019+  0.040+    0.051}$ & $    0.840$\\
			$H_0$ & $-$ & $   68.95_{- 0.95-    1.78-    2.22}^{+    0.93+    1.84+    2.34}$ & $   68.13$\\
			${\rm{Age}}/{\rm{Gyr}}$ & $-$ & $   13.803_{-    0.046-    0.090-  0.113}^{+ 0.047+    0.089+ 0.121}$ & $   13.817$\\
			\hline\hline
		\end{tabular}
		\caption{The table summarizes the mean values with $1\sigma$ (68.3\%), $2\sigma$ (95.5\%), and $3\sigma$ (99.7\%) confidence -level uncertainties of the cosmological parameters for the interacting dark energy model using the observational data CC $+$ $H_0$ $+$ JLA $+$ BAO $+$ CMB (Planck TT $+$ lowP) where $\Omega_{m0} = \Omega_{dm,0}+ \Omega_{b0}$. }
		\label{tab:results}
	\end{table}
\end{center}

\begin{figure*}[!htbp]
	\includegraphics[width=18cm,height=16cm]{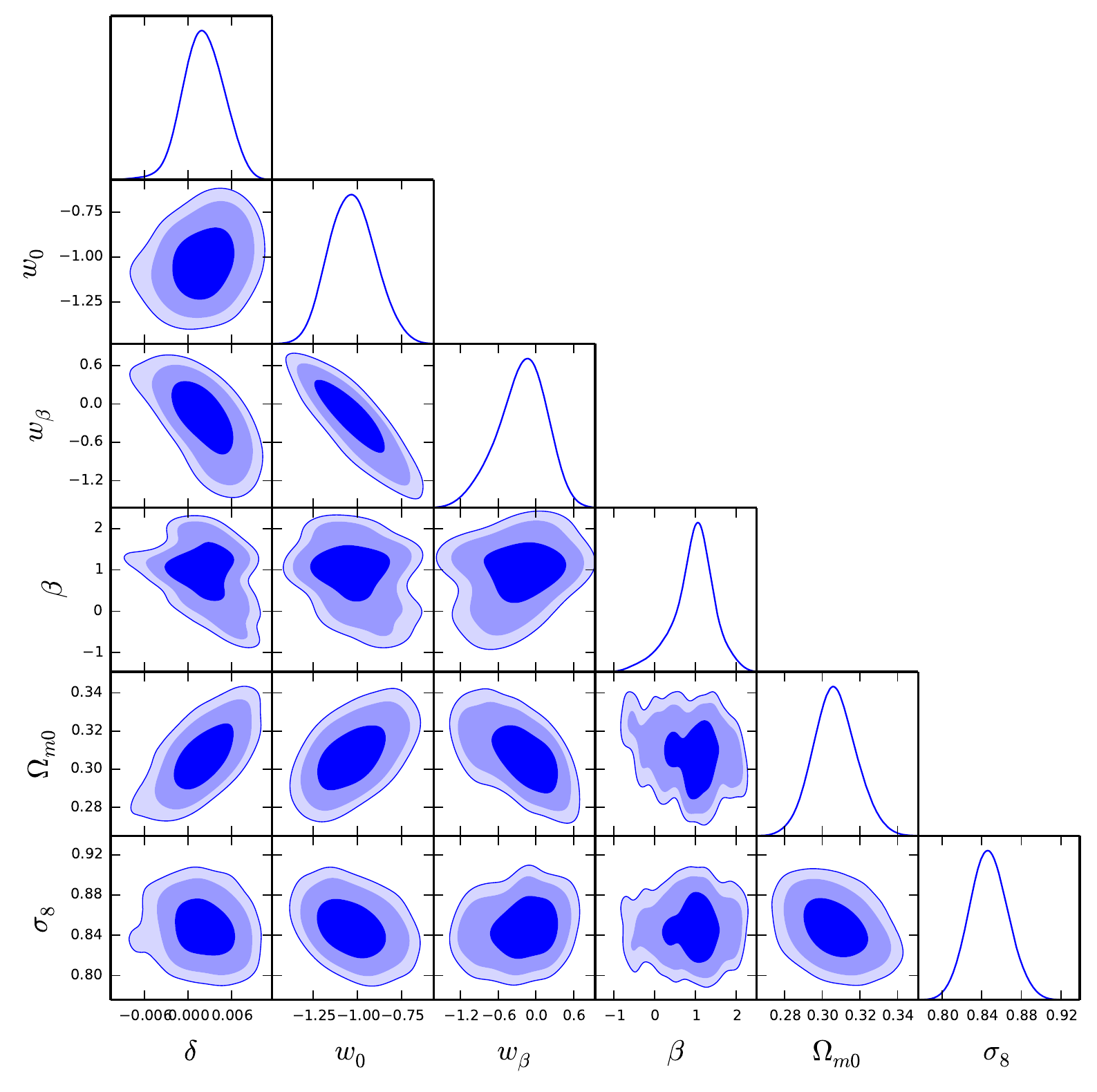}
	\caption{68.3\%, 95.5\%, and 99.7\% confidence-level contour plots for the various pairs of the free parameters of the interacting scenario have been shown using the observational data CC $+$ $H_0$ $+$ JLA $+$ BAO $+$ CMB (Planck TT $+$ lowP). Additionally, we have also shown the 1-dimensional marginalized posterior distributions of individual free parameters. We note that $\Omega_{m0}= \Omega_{dm,0}+ \Omega_{b0}$. }
	\label{fig:contour}
\end{figure*}

\subsection{Other standard dark energy probes}

\subsubsection{Type Ia Supernovae}
\label{snia-data}

Supernovae Type Ia (SNIa) provided the first signal for an accelerating universe, and still they serve as the main observational data to probe the late-time acceleration of the universe. In this work we consider the joint light curves (JLA) sample \cite{ref:JLA} containing 740 SNIa data in the redshift range $z\in[0.01, 1.30]$.

\subsubsection{Baryon Acoustic Oscillation distance measurements}
\label{bao-data}

For the Baryon acoustic oscillation (BAO) distance measurement  data set, we use the measured ratio of $r_s/D_V$ as a 'standard ruler', where $r_s$ is the comoving sound horizon at the baryon drag epoch. $D_V$ is the effective distance which is determined by the angular diameter distance $D_A$ and Hubble parameter $H$ through the relation $D_V(z)=\left[(1+z)^2D_A(a)^2\frac{z}{H(z)}\right]^{1/3}$. At three different redshifts, $r_s(z_d)/D_V(z=0.106)=0.336\pm0.015$ from 6-degree Field Galaxy Redshift Survey (6dFGRS) data \cite{ref:BAO-1}, $r_s(z_d)/D_V(z=0.35)=0.1126\pm0.0022$ from Sloan Digital Sky Survey Data Release 7 (SDSS DR7) data \cite{ref:BAO-2}, and $r_s(z_d)/D_V(z=0.57)=0.0732\pm0.0012$  from SDSS DR9 \cite{ref:BAO-3} are utilized. \\
We pick up only this data for BAO so as to minimize the use of data sets which are correlated. We refer to the Ref. \cite{Ade:2013zuv} for some supportive argument.

\subsubsection{Cosmic Microwace Background data}

The cosmic microwave background (CMB) data from Planck 2015 measurements  \cite{ref:Planck2015-1,ref:Planck2015-2} have been used in our analysis. Here, we combine the likelihood of full Planck temperature-only $C^{TT}_l$ with the low$-l$ polarization $C^{TE}_l+C^{EE}_l+C^{BB}_l$, which in notation is  same with ``PlanckTT $+$ lowP'' of Ref. \cite{Ade:2015xua}.

\section{Cosmological constraint results}
\label{sec:results}

The total likelihood for our analysis follows $\mathcal{L}\propto e^{-\chi^2_{tot}/2}$, where $\chi^2_{tot}$ is given by
\begin{eqnarray}
\chi^2_{tot}=\chi^2_{CC}+\chi^2_{H_0}+\chi^2_{JLA}+\chi^2_{BAO}+ \chi^2_{CMB}.
\label{eq:chi22}
\end{eqnarray}
For the present interacting dark energy (IDE) model, we modify the publicly available code CAMB \cite{cosmomc} where we add a numerical algorithm. Then we call this numerical algorithm in order to solve the background equations and for each data set we calculate the corresponding $\chi^2$ values. Finally we call CosmoMC, a markov chain monte carlo simulation method to explore the cosmological parameter space. Here we have the
following ten-dimensional parameter space
\begin{align}
P\equiv\Bigl\{\Omega_bh^2, \Omega_{dm}h^2, \Theta_S, \tau, \delta, w_0, w_{\beta}, \beta, n_s, log[10^{10}A_S]\Bigr\},
\label{eq:parameter_space}
\end{align}
where $\Omega_bh^2$ and $\Omega_{dm}h^2$, respectively, stand for the density of the baryons and dark matter, $\Theta_S=100\theta_{MC}$ refers to the ratio of sound horizon and angular diameter distance, $\tau$ indicates the optical depth, $\delta, w_0, w_{\beta}, \beta$ are the characteristic parameters of the IDE model, $n_s$ is the scalar spectral index, and $A_s$ represents the amplitude of the initial power spectrum.

In Table \ref{tab:results}, we summarize the results  of the current interacting scenario using the latest observational data  CC $+$ $H_0$ $+$ JLA $+$ BAO $+$ CMB (Planck TT $+$ lowP) while in Fig. \ref{fig:contour} we show the $68.4\%$, $95.5\%$, and $99.7\%$ confidence-level contour plots for different pairs of free parameters of the interacting scenario as well as the one dimensional marginalized distribution of individual free parameters. In the contour plots, along with $\delta , w_{0}, \beta , w_{\beta}$ and ${\Omega}_{m0}$, we also take into account ${\sigma}_{8}$ which is a measure of the amplitude of the linear power spectrum on the scale of $8h^{-1}$Mpc. From our analysis, we find that the interaction term $\delta$ is compatible with zero within the 1$\sigma$ confidence limit.
On the other hand, from Table \ref{tab:results}, it is clearly seen that although the mean value of $w_0$ goes beyond the `$-1$' boundary but its best fit value still represents a quintessential dark energy in presence of the interaction between the dark sectors. Further, one may notice that the interacting model can relieve the tension on $H_0$ as observed independently by the recent Planck's estimation ($H_0= 67.27 \pm 0.66$ km/s/Mpc) \cite{Ade:2015xua} and the local measurements by Riess et al ($H_0= 73.24 \pm 1.74$ km/s/Mpc) \cite{Riess:2016jrr} within the $\Lambda$CDM framework. This tension can be relieved at about $2\sigma$ confidence level in the direction of Planck's measurements \cite{Ade:2015xua}. However, one may expect this decrease in tension due to enlargement of the parameter space.  In the following subsection we shall present some other cosmological consequences that are directly related to the interaction parameter, $\delta$.


\begin{figure}[!htbp]
	\includegraphics[width=9cm,height=7cm]{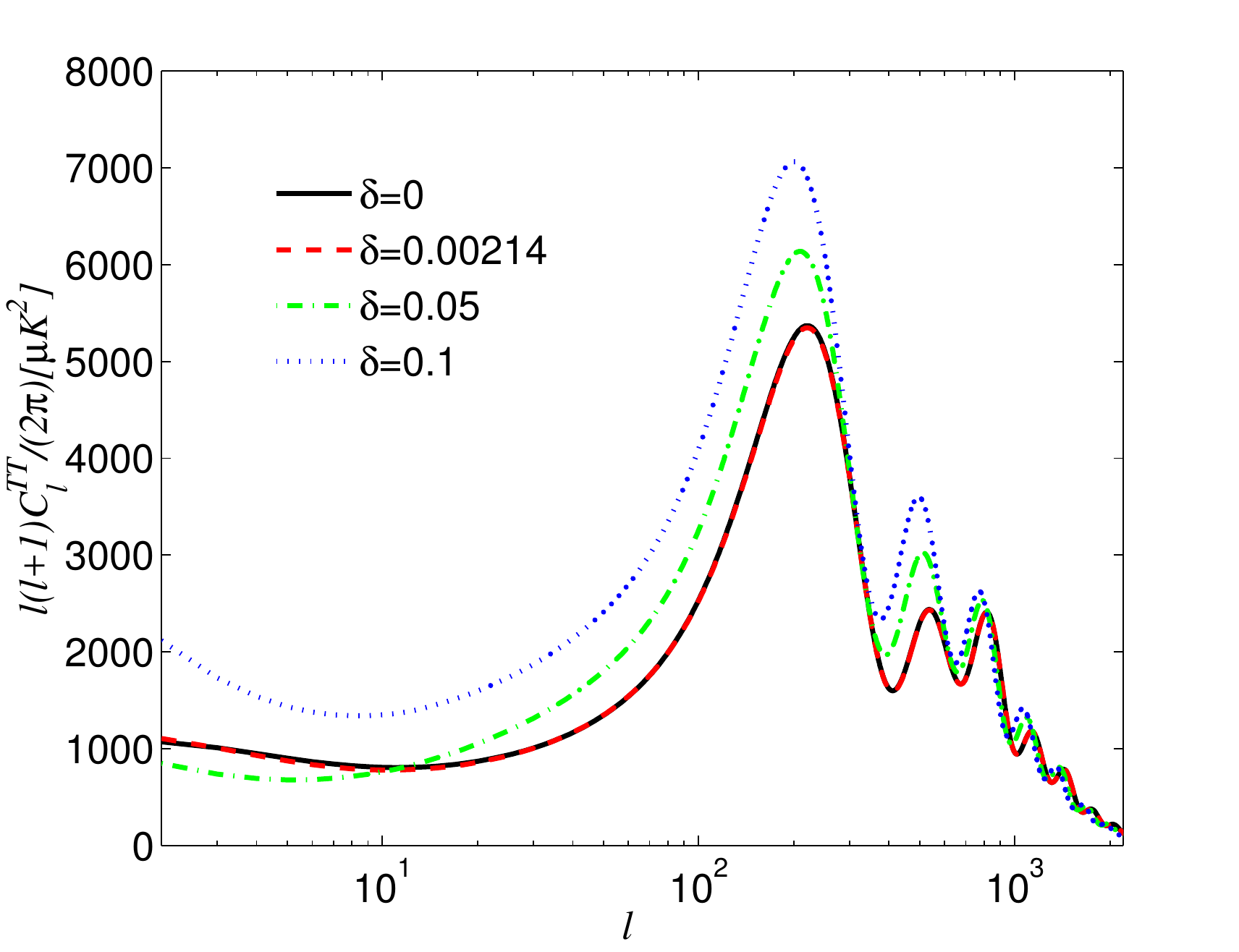}
	\caption{The effects on the CMB temperature power spectra for different values of the interaction parameter $\delta$. The black solid, red thick dashed, green dash-dot, and blue dotted lines are for $\delta=0, 0.00214, 0.05$, and $0.1$, respectively; the other relevant parameters are fixed with the mean values as shown in the third column of Table \ref{tab:results}. We note that the curves for $\delta=0$ (black solid) and $\delta = 0.00214$ (red thick dashed) are almost indistinguishable. }
	\label{fig:CMBTTpower}
\end{figure}

\begin{figure}[!htbp]
	\includegraphics[width=8.5cm,height=6.5cm]{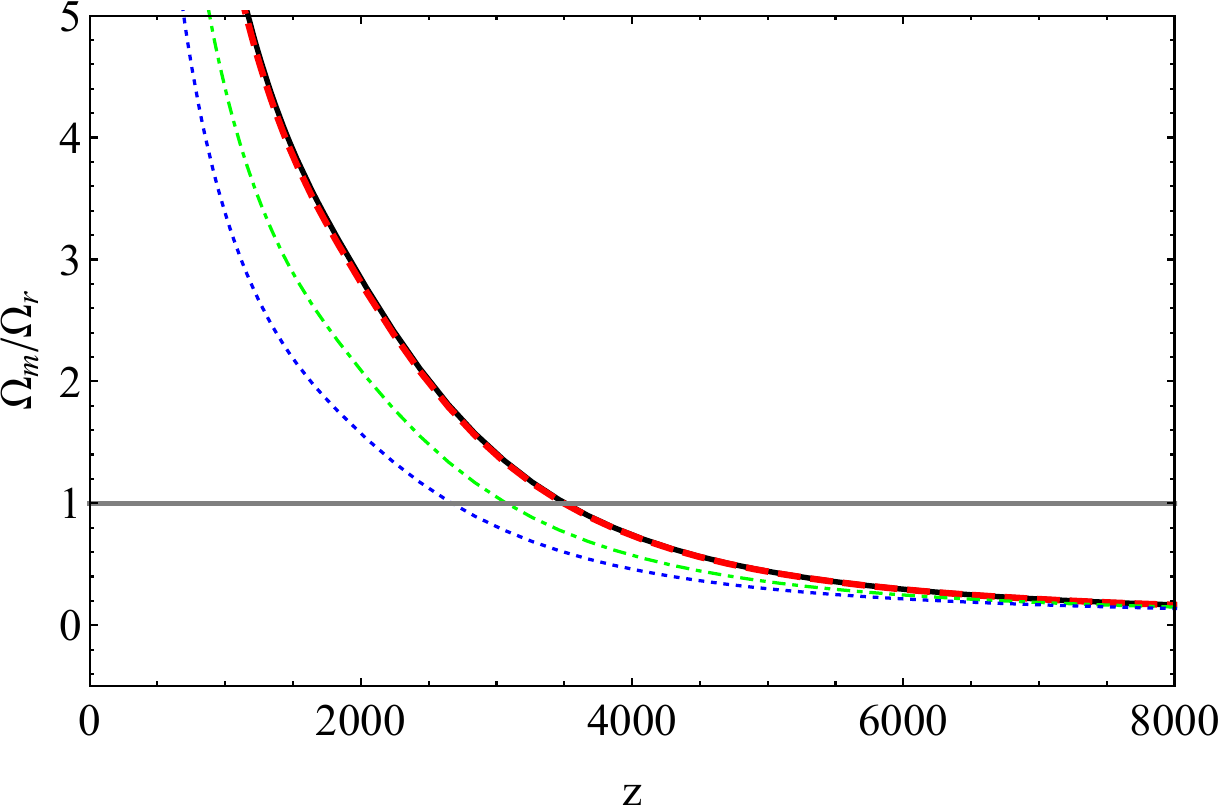}
	\caption{The evolution for the ratio of matter and radiation $\Omega_m/\Omega_r$ (Here, $\Omega_m= \Omega_{dm}+ \Omega_b$) when the interaction parameter $\delta$ is varied. The different lines correspond to the cases of the Fig. \ref{fig:CMBTTpower}; the horizontal gray thick line responds to the case of $\Omega_m=\Omega_r$, and the other relevant parameters are fixed with the mean values as shown in the third column of Table \ref{tab:results}. We see that the curves for $\delta=0$ (black solid) and $\delta = 0.00214$ (red thick dashed) are almost indistinguishable}
	\label{fig:OmOr}
\end{figure}

\begin{figure}[!htbp]
	\includegraphics[width=9cm,height=7cm]{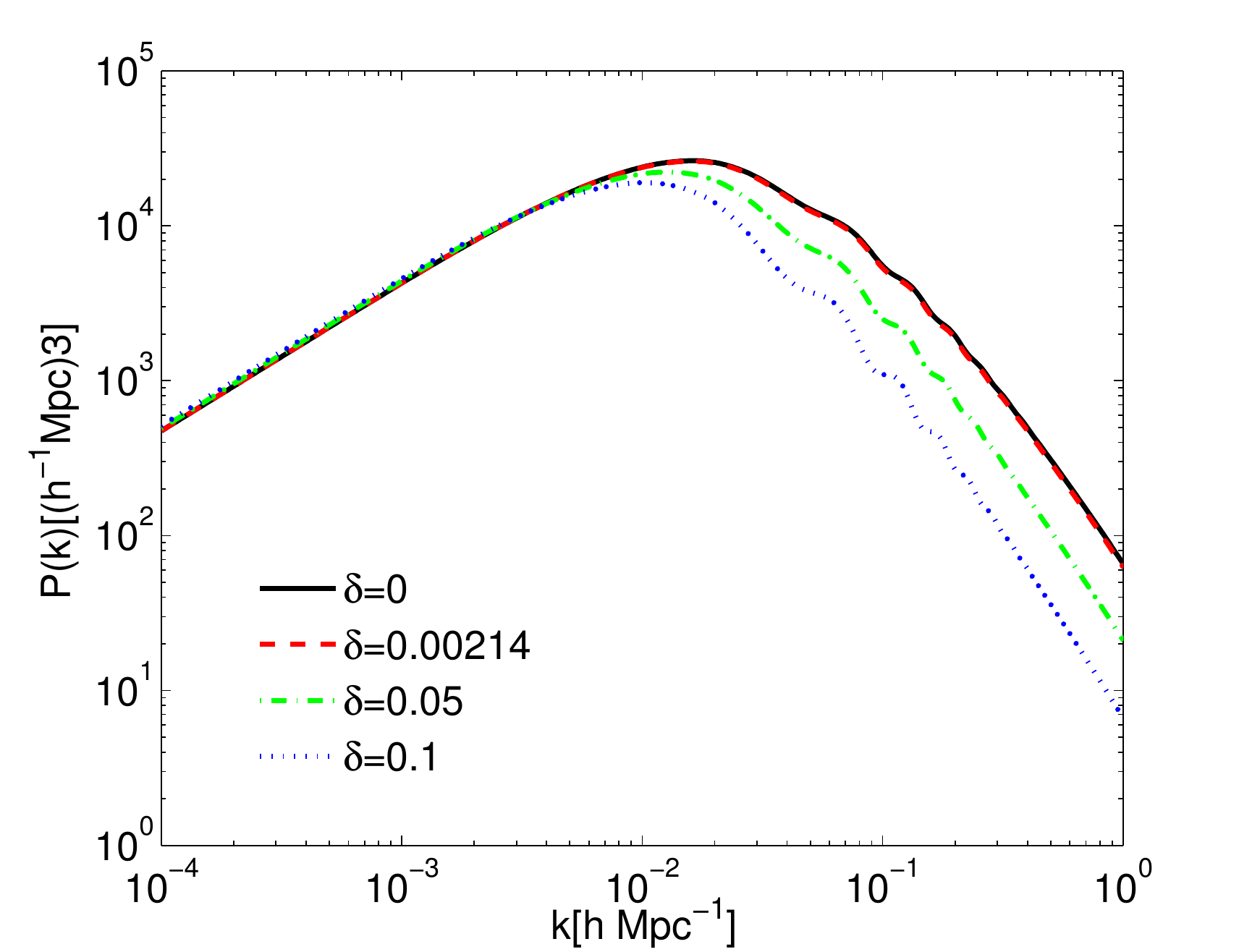}
	\caption{The effects on the matter power spectra for different values of the interaction parameter $\delta$. The black solid, red thick dashed, green dotted-dashed, and blue dotted lines are for $\delta=0, 0.00214, 0.05$, and $0.1$, respectively; the other relevant parameters are fixed with the mean values as shown in the third column of Table \ref{tab:results}. Similar to the Figures \ref{fig:CMBTTpower} and \ref{fig:OmOr} we find that the curves for $\delta=0$ (black solid) and $\delta = 0.00214$ (red thick dashed) cannot be distinguished from each other. }
	\label{fig:Mpower}
\end{figure}

\begingroup

\begin{center}                                                                                  \begin{table}                                                                                               \begin{tabular}{cccccccc}                                                    \hline\hline                                                                                                  Parameters & Priors & Mean with errors & Best fit \\ \hline
			
			$\Omega_b h^2$ & $[0.05, 0.1]$ & $    0.02213_{-    0.00026-    0.00044-    0.00055}^{+    0.00021+    0.00046+    0.00063}$ & $    0.02227$\\
			$\Omega_{dm} h^2$ &  $[0.01, 0.99]$ & $    0.1189_{-    0.0016-    0.0035-    0.0048}^{+    0.0020+    0.0030+    0.0044}$ & $    0.1177$\\
			$100\theta_{MC}$ & $[0.5, 10]$ & $    1.04094_{-    0.00041-  0.00093-    0.00110}^{+    0.00050+    0.00080+    0.00104}$ & $    1.04086$\\
			$\tau$ &  $[0.01, 0.8]$ &$    0.069_{-    0.020-    0.033-    0.042}^{+    0.017+    0.038+    0.054}$ & $    0.089$\\
			$n_s$ & $[0.5, 1.5]$ & $    0.9639_{-    0.0060-    0.0097-    0.0128}^{+    0.0042+    0.0107+    0.0157}$ & $    0.9703$\\
			${\rm{ln}}(10^{10} A_s)$ & $[2.4, 4]$ & $    3.076_{-    0.038-    0.065-    0.084}^{+    0.034+    0.070+    0.108}$ & $    3.113$\\
			$\beta$ &  $[-3, 3]$ & $    0.959_{- 0.071- 0.154- 0.177}^{+ 0.068+    0.170+    0.212}$ & $    0.928$\\
			$w_0$ & $[-2, 0]$ & $   -1.037_{-    0.045-    0.087-    0.135}^{+    0.045+    0.092+    0.118}$ & $   -1.095$\\
			$w_{\beta}$ & $[-3, 3]$ & $    0.022_{-    0.085- 0.167- 0.191}^{+ 0.067+  0.195+ 0.223}$ & $    0.204$\\
			$\eta$ & $[0, 1]$ & $    0.00062_{- 0.00058-    0.00062- 0.00062}^{+    0.00021+ 0.00081+  0.00107}$ & $    0.00061$\\
			$\Omega_{m0}$ & $-$ & $    0.298_{- 0.010- 0.017- 0.024}^{+ 0.009+  0.017+ 0.025}$ & $    0.301$\\
			$\sigma_8$ &  $-$ & $ 0.829_{- 0.017-    0.037- 0.047}^{+ 0.016+ 0.035+ 0.047}$ & $    0.854$\\
			$H_0$ & $-$ & $   68.92_{-  1.06-    1.76-    2.50}^{+  0.99+    1.81+    2.49}$ & $   68.39$\\
			${\rm{Age}}/{\rm{Gyr}}$ & $-$ & $   13.747_{-    0.037- 0.066 - 0.082}^{+    0.035+    0.064 + 0.087}$ & $   13.783$\\
			\hline\hline                                                                                               \end{tabular}                                          \caption{The table summarizes the mean values with their errors at $1\sigma$ (68.3\%), $2\sigma$ (95.5\%), and $3\sigma$ (99.7\%) confidence-level errors of the non-interacting scenario (\ref{noninteraction}) presented in Section \ref{pressure} for the observational data CC $+$ $H_0$ $+$ JLA $+$ BAO $+$ CMB (Planck TT $+$ lowP) where $\Omega_{m0}= \Omega_{dm,0}+ \Omega_{b0}$. }
		\label{table-noninteraction}                
		\end{table}                                                                                 \end{center}                                                                      \endgroup

\begin{figure*}[!htbp]
	\includegraphics[width=18cm,height=16cm]{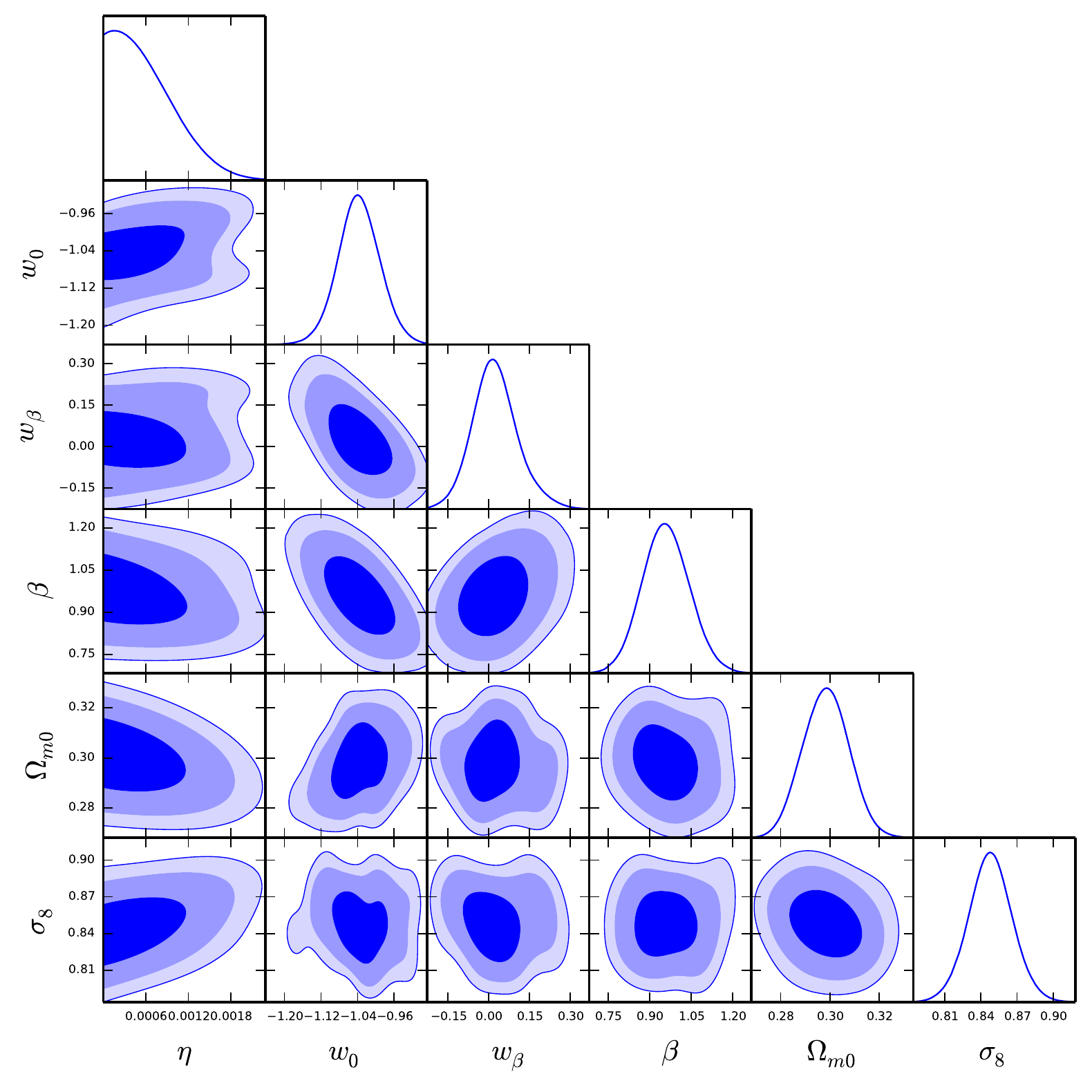}
	\caption{68.3\%, 95.5\%, and 99.7\% confidence-level contour plots for the various pairs of the free parameters of the noninteracting scenario (eqn. (\ref{noninteraction})) have been shown using the observational data CC $+$ $H_0$ $+$ JLA $+$ BAO $+$ CMB (Planck TT $+$ lowP). Additionally, we have also shown the 1-dimensional marginalized posterior distributions of individual free parameters. Here, let us note that $\Omega_{m0}= \Omega_{dm,0}+ \Omega_{b0}$. }
	\label{fig:contour2}
\end{figure*}


\subsection{Cosmological Implications: Theoretical predictions of CMB temperature and matter power spectra}

\label{sec-temperature}

The interaction in the dark sectors can produce some effects, for instance in the CMB temperature power spectra as well as in the matter power spectra.  In Figure \ref{fig:CMBTTpower}, the influences on the CMB temperature power spectra have been shown for different interaction parameter $\delta$. Further, in order to show the relation between the interaction parameter $\delta$ and the moment of matter-radiation equality, we also plot the evolution curves of $\Omega_m/\Omega_r$ in Figure \ref{fig:OmOr} which shows that
the increasing in the interaction parameter $\delta$ results in a decreasing density parameter of the effective matter $\Omega_m$, which also delays the moment of matter-radiation equality and hence, the sound horizon is increased. As a result, the first peak of CMB temperature power spectra is enhanced for higher values of $\delta$ as seen in Figure \ref{fig:CMBTTpower}.
Further, at large scales ($l<100$), the integrated Sachs-Wolfe (ISW) effect becomes dominant, the parameter $\delta$ quantifying the interaction rate, affects the CMB power spectra via ISW effect due to the evolution of gravitational potential. In Fig. \ref{fig:Mpower}, we plot the influence of the interaction on the matter power spectrum $P(k)$. We find that the evolution is just the opposite of the CMB temperature power spectra depicted in Figure  \ref{fig:CMBTTpower}. Now, with the increasing values of $\delta$, meaning a stronger interaction, the matter power spectra $P(k)$ are depressed due to the late matter-radiation equality. The case of $\delta=0.00214$ (which corresponds to the interacting scenario with the mean value of $\delta$) and that of $\delta=0$ (corresponds to the uncoupled $w$CDM model) are almost same as depicted from the latest observational data.

\section{An interation between DE and DM or a small pressure ?}
\label{pressure}

So long as $\delta$ is a constant, the evolution ansatz $\rho_{dm} = \rho_{dm,0}\, a^{-3+\delta}$ as in equation (\ref{cdm-evolution}) clearly indicates that this can be achieved by means of an equation of state $p= \eta \rho$, (where $\delta = -3\eta$) without any interaction, i.e., with $Q=0$. This $\delta$, with its new connotation, can now be constrained by the observational data. With the same equation of state for the dark energy as in (\ref{model1}), the Hubble function can be written as

\begin{align}\label{noninteraction}
	\left(\frac{H}{H_0}\right)^2 = \Omega_{r0}(1+z)^4+ \Omega_{b0}(1+z)^3+ \Omega_{dm,0}(1+z)^{3+ 3 \eta}\nonumber\\ + \Omega_{DE,0} (1+z)^{3 \Bigl(1+w_0 + \frac{w_{\beta}}{\beta} \Bigr)}\, \exp\left( \frac{3 w_{\beta}}{\beta^2} (1+z)^{-\beta} \right).
\end{align}

With the same statistical analysis, we find an interesting result that it is indeed possible to have small positive value of $\eta$, consistent with the observations. The most likely value of $\eta$ comes out to be ``$0.00062$'' indicationg a very small pressure for the dark matter. The summary of the mean values as well as the best-fit values of various parameters are given in Table \ref{table-noninteraction}. Figure \ref{fig:contour2} contains 2D contours between various pairs of parameters and also 1D marginalized distribution of individual parameters. \\

The motivation behind this section is to show that the same evolution of the dark matter density as well as nearly identical observational quantities can be given by an interaction in the dark sector or eqivalently by means of a small pressure of the dark matter.



\section{Summary and Discussion}
\label{sec:summary}

In this work we have considered an interaction between the dark energy and the cold dark matter where the rate of interaction is measured by a quantity `$\delta$', considered to be very slowly varying. The equation of state parameter $w_{DE}$ for the  dark energy has been considered to be a variable as a general case. The variable EoS in DE has been taken as in (\ref{model1}) characterized by a  parameter $\beta$ which recovers some well known and frequently used DE parametrizations, such as CPL (eqn. (\ref{w-cpl})), linear (see eq. (\ref{w-linear})), and logarithmic (see eqn. (\ref{w-log})). Now, using some recent   data sets, namely, (i) the cosmic chronometers, (ii) local value of the Hubble parameter, (iii) Joint light curves of Supernovae Type Ia, (iv) baryon acoustic oscillations distance measurements and finally (v) the cosmic micorwave background, we have constrained the whole interacting scneario with the help of Markov Chain Monte Carlo (MCMC) algorithms. The results have been summarized in Table \ref{tab:results} and the contour plots for the free parameters have been shown in Figure \ref{fig:contour}. From our analysis, we find that no interaction is compatible with observations within the 1$\sigma$ confidence limit. This result is in agreement with some recent investigations where the EoS of DE was kept to be a constant \cite{Nunes:2016dlj, Kumar:2016zpg}. \\

It is noted that the interacting model could also alleviate the current tension at about $2\sigma$ confidence level in the $\Lambda$CDM based Hubble constant measurements by Planck collaborations \cite{Ade:2015xua}. However, this compatibility could be due to the enlargement in the parameter space.\\

One can see that the interaction parameter $\delta$ plays a  role in some other sectors in cosmology. It is evident  that the increasing value of the interaction parameter, i.e., $\delta$, enhances the first peak in the CMB temperature power spectra (see Figure \ref{fig:CMBTTpower}); delays the matter-radiation equality (see Figure \ref{fig:OmOr}); and finally makes a clear deviation in the matter power spectra for large $l < 100$ (see Figure \ref{fig:Mpower}). \\

If we pretend that there is no interaction in the dark sector, but there is rather a tiny pressure of the CDM, so that the evolution of the dark matter density is a bit different from $a^{-3}$ and is given by $a^{-3 +\delta}$, we see that the parameters constrained by observations are hardly different for the interacting scenario (see Tables I and II)! The major difference is that $\delta = -3\eta$ is negative (as $\eta > 0$), so the matter density redshifts faster than $a^{-3}$ as opposed to the interacting scenario where ${\rho}_{m}$ decays at rate slower than  $a^{-3}$. This description has an advantage that as there is no flow of energy from one dark sector to another, there is no thermodynamic compulsion on the signature of $\eta$ which is the constant equation of state parameter in this case. In fact a positive value is welcome as it does not invoke any exotic matter. This idea of reproducing the departure from the standard evolution of $\rho_{dm}$ through a small pressure rather than an interaction is indeed new, but perhaps not unphysical. The reason for assuming $p=0$ for the CDM is that the energy is solely the rest energy. A tiny pressure would indicate that the energy is largely rest energy, but the kinetic energy, though much smaller, has a small contribution.\\

Tables I and II clearly indicate that the present value of the equation of state parameter, $w_{DE}$, is very close to the phantom bound of `$-1$'. In fact, in the non-interacting case, i.e., when the particular evolution ansatz of the dark matter density is achieved via a small pressure, both the mean and the best-fit for $w_{DE}$ are beyond the phantom divide ($<-1$). For the interacting case, however, the best-fit value for $w_{DE}$ is marginally non-phantom ($> -1$). The nature of dark energy in this interacting scenario thus resembles that with a constant $w_{DE}$ in the presence of a background of neutrino distribution \cite{Kumar:2016zpg} but differs from that without any neutrino background \cite{Nunes:2016dlj}.\\

\bigskip

\section*{Acknowledgments}
The authors thank the referee for some clarifying comments that were helpful to improve the mansucript. W. Y.'s work is supported by the National Natural Science Foundation of China under Grant No. 11647153, the Foundation of Education Department of Liaoning Province in China under Grant No. L201683666, and the Youth Foundation of Liaoning Normal University under Grant No. LS2015L003. SP is supported by the SERB-NPDF grant (File No: PDF/2015/000640), Government of India.

\end{document}